\documentclass[pre,twocolumn,superscriptaddress,nofootinbib]{revtex4}

\usepackage{graphicx,chemarr}

\begin{document}

\title{Spiral wave propagation in communities with spatially correlated heterogeneity}

\author{Xiaoling Zhai}
\affiliation{Department of Physics and Astronomy, Purdue University, West Lafayette, Indiana 47907, USA}

\author{Joseph W.\ Larkin}
\affiliation{Division of Biological Sciences, University of California San Diego, Pacific Hall Room 2225B, Mail Code 0347, 9500 Gilman Drive, La Jolla, CA 92093, USA}

\author{G\"urol M.\ S\"uel}
\affiliation{Division of Biological Sciences, University of California San Diego, Pacific Hall Room 2225B, Mail Code 0347, 9500 Gilman Drive, La Jolla, CA 92093, USA}
\affiliation{San Diego Center for Systems Biology, University of California San Diego, La Jolla, CA 92093, USA}

\author{Andrew Mugler}
\email{amugler@purdue.edu}
\affiliation{Department of Physics and Astronomy, Purdue University, West Lafayette, Indiana 47907, USA}

\begin{abstract}
Many multicellular communities propagate signals in a directed manner via excitable waves. Cell-to-cell heterogeneity is a ubiquitous feature of multicellular communities, but the effects of heterogeneity on wave propagation are still unclear. Here we use a minimal FitHugh-Nagumo-type model to investigate excitable wave propagation in a two-dimensional heterogeneous community. The model shows three dynamic regimes in which waves either propagate directionally, die out, or spiral indefinitely, and we characterize how these regimes depend on the heterogeneity parameters. We find that in some parameter regimes, spatial correlations in the heterogeneity enhance directional propagation and suppress spiraling. However, in other regimes, spatial correlations promote spiraling, a surprising feature that we explain by demonstrating that these spirals form by a second, distinct mechanism. Finally, we characterize the dependence of the spiral period on the degree of heterogeneity in the system by using techniques from percolation theory. Our results reveal that the spatial structure of cell-to-cell heterogeneity can have important consequences for signal propagation in cellular communities.\end{abstract}

\maketitle

\section{Introduction}
Collective behaviors are omnipresent in natural systems. They exist in diverse systems such as animal flocking, microbial colony formation and synchronization, traffic jamming, and social segregation in human populations \cite{BalleriniPNAS,JacobAP,FlynnPRE,DaninoNature,SchellingJMS}. Many of these systems span a hierarchy of scales and complexity, but the collective behavior at larger scales can often be understood without knowledge of many details at smaller scales \cite{NoorbakhshPRE,SchellersheimWIR,QutubEMBM,SouthernPBMB}. This coarse-grained approach is often used to describe communities where a directed signal travels from agent to agent to serve a function, which permits connections to reaction-diffusion systems of excitable media \cite{LilienkampPRL}.

Directed signal propagation in multicellular communities is an important and commonly observed phenomenon. In the nervous system, a neuron passes an electrochemical signal to another neuron directly through a synapse. In a biofilm of {\it Bacillus subtilis} bacteria, the interior cells transmit an electrochemical wave to the periphery to mediate the metabolic activity \cite{PrindleNature,LiuNature}. In pancreatic islets, gap junction coupling mediates electrical communication between cells \cite{BenningerBioPhysJ}. The social amoebae {\it Dictyostelium discoideum} communicate via the signaling molecule cyclic adenosine monophosphate (cAMP) to coordinate aggregation \cite{NoorbakhshPRE}. 

Heterogeneity in these communities may pose a challenge to directed wave propagation. The stochastic expression of genes leads to non-uniform protein distributions in cells. As a result, cells have heterogeneous responses to the wave propagation, and obstacles are naturally generated. For example, in the biofilm, a fraction of cells do not participate in the signal transmission, which can cause propagation failure \cite{JosephCellSystem, zhai2019statistics}. In cardiac cells, cell death can result in large patches of non-conductive tissue \cite{BubChaos, KinoshitaPRE}, which are related to heart disease such as fibrosis and ischemia \cite{BubChaos, UzzamanCR}. Heterogeneities play an important role in deciding the wave pattern and result in a variety of wave behaviors, including wave death and spirals \cite{KinoshitaPRE}. However, the impact of heterogeneity on wave propagation is still not fully understood. 

The influence of heterogeneity on wave propagation has been investigated experimentally using cell monolayers \cite{BubPRL} and patterned chemical systems \cite{SteinbockScience, TinsleyPCCP, TothPRE}, and theoretically using the FitzHugh-Nagumo model and cellular automata models \cite{BubChaos}. When the length scales of the heterogeneities in the medium are close to the length scales associated with the propagating front, propagating waves can develop breaks, which can either block wave propagation or cause spiraling \cite{BubChaos}.
There are several mechanisms to explain the formation of spirals. For example the heterogeneity of the reaction field can stochastically generate unidirectional sites, which can induce spirals \cite{KinoshitaPRE}. Spiral waves can also form as a result of an interaction between chemical waves \cite{GesteiraPD,AlievCSF}: when a chemical wave comes close to another wave from the back, part of the wave vanishes because of the refractory region of the preceding wave, and as a result spiral cores are formed.

Spiral waves can be either beneficial or detrimental to function. For example, spiral reentry of the action potential can induce heart failure \cite{PertsovCirR, ZimikArXiv}, whereas a spiral pattern formed on the nest of honeybees is an effective protection mechanism \cite{KastbergerPO}. Therefore it is important to understand the formation of spiral waves in order to better understand community-level function in living systems. 

Here we focus on excitable wave propagation in a two-dimensional heterogeneous community. We use a minimal FitzHugh-Nagumo-type model to describe the excitability of cells and introduce heterogeneity in the model parameters. Our model predicts three dynamic regimes (directed propagation, wave death, and spiraling), and we demonstrate how these regimes depend on the heterogeneity. We then introduce spatial correlations in the heterogeneity, which naturally arise due to lineage-based inheritance and cell-to-cell interactions \cite{zhai2019statistics}. Intuitively, one expects correlations to cause channeling and therefore promote directed propagation, and we indeed find this result in particular parameter regimes. However, we also find in other parameter regimes that correlations promote the emergence of spiral waves, a surprising feature that we explain by distinguishing between two different spiraling mechanisms. Finally, we characterize the dependence of the spiral period on the degree of heterogeneity in the system using percolation theory. Our results suggest that the spatial structure of cell-to-cell heterogeneity can have important consequences for signal propagation in cellular communities.

\section{Materials and methods}
We construct an integrated model combining the FitzHugh-Nagumo (FN) model \cite{FitzHugh,Nagumo} and spatial heterogeneity. We use the FN model to describe the excitation of a single cell, which is coupled to its four neighbors in a square lattice. Heterogeneities are introduced by randomly assigning each cell to be ``on'' or ``off'' which determines its parameter values (Fig.\ \ref{Fig1}), as described below.

\subsection{Excitable wave model}
Hodgkin and Huxley proposed the first quantitative mathematical model to describe action potential propagation through very careful experimentation on the squid giant axon \cite{HodgkinJP}. Modifying the Hodgkin and Huxley model for excitable media, FitzHugh and Nagumo established a partial differential equation model of two variables \cite{FitzHugh,Nagumo}. The FN model is a simplified variant of the Hodgkin and Huxley model, which traces the fast-slow dynamics of an excitable system \cite{OsmanISRN}.

We use a minimal FitHugh-Nagumo-type model \cite{Tuckwell} to describef excitable wave propagation in a two-dimensional heterogeneous community. This minimal model is
\begin{align}
\label{rd1}
&\frac{du_i}{dt}=\epsilon [u_i^2 (1-u_i) -w_i]+ \sum_{j\in {\cal N}(i)} (u_j-u_i),\\
\label{rd2}
&\frac{dw_i}{dt}=\frac{1}{\tau} u_i,
\end{align}
where $i$ indexes the cells, and ${\cal N}(i)$ denotes the neighbors of cell $i$. The fast activator variable $u$ and the slow inhibitor variable $w$ are known as the excitable and recovery variable, respectively. The variables $u$ and $v$ are also referred to as the propagator and controller variables, respectively. The parameters in this reaction-diffusion system are the excitation strength $\epsilon$, which when sufficiently high supports pulse-coupled wave propagation \cite{Mirollo}, and the recovery time $\tau$, which sets the pulse duration of the signal and thus governs the dynamics. The second term in Eq.\ \ref{rd1} is a discretized Laplacian term to account for the cell-cell communication. The minimalistic nature of this model promotes the interpretation of dynamics for single cells in excitable media.

\begin{figure}
	\center{\includegraphics[width=0.5\textwidth]
		{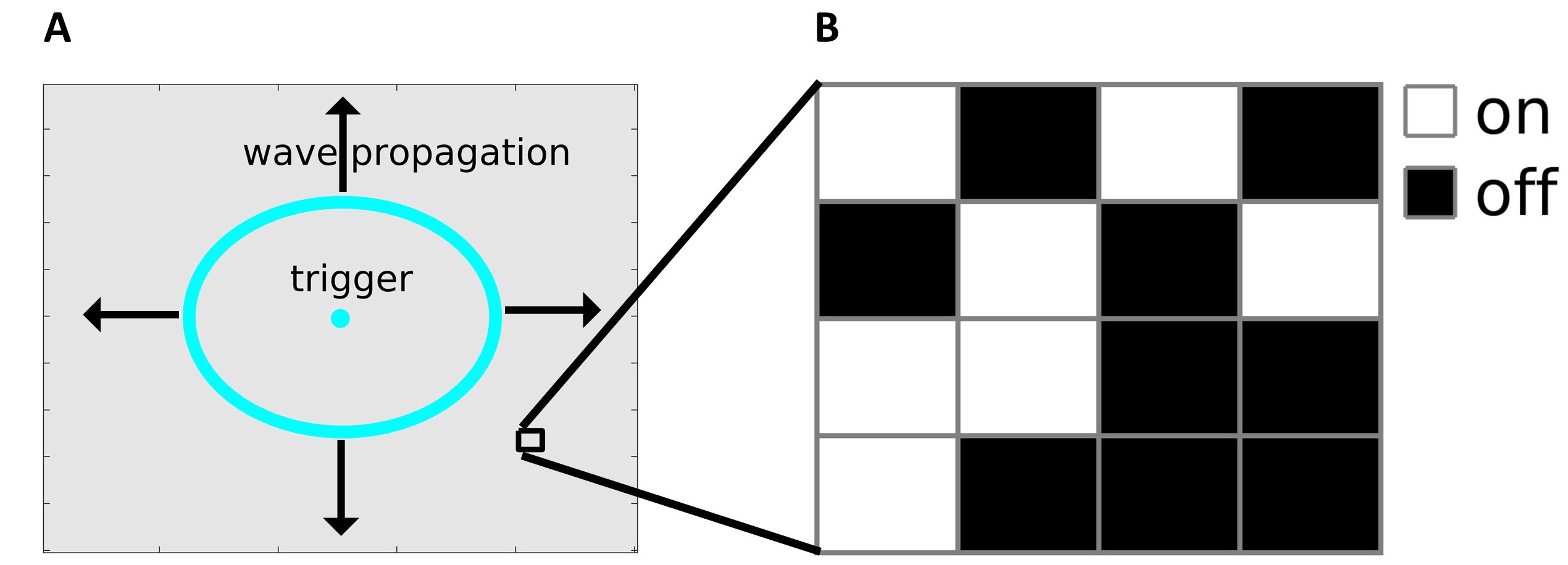}  }
	\caption{Excitable wave propagation model in a two-dimensional heterogenous medium. (A) We evolve the dynamics on a $100\times100$ square lattice. A wave is triggered by the central 5 by 5 cells and propagates outward to the edges (cyan). (B) To introduce the heterogeneity, each cell is randomly assigned to be ``on'' (small $\tau$) or ``off'' (large $\tau$).} \label{Fig1}
\end{figure}

In all dynamical simulations, cells in the center are initialized with $u = 1$ to trigger the excitable wave; all other cells are initialized with $w = 0$ (Fig.\ \ref{Fig1}A). We use a square lattice of $100\times100$ cells with absorbing boundaries on four sides. To evolve the dynamics, we discretize the FN model in time using the fourth-order Runge-Kutta method with time step $\Delta t$ = 0.02.

\subsection{Introduction of heterogeneity}
We start with a homogeneous community (all cells have the same parameter values), and set the parameters in the following way to support directed propagation. The excitation strength $\epsilon$ must be larger than 1 because otherwise diffusion outpaces excitation and washes away the signal; therefore we set $\epsilon=10$ \cite{JosephCellSystem}. This places the model in a diffusion-limited regime, which is consistent with signaling in the biofilms of {\it Bacillus subtilis} \cite{PrindleNature}. We find that the recovery time $\tau$ needs to be sufficiently large ($\tau > 7$) because otherwise the excitation is too short-lived to trigger its neighbors with $\epsilon=10$. 

To introduce heterogeneity, we must choose (i) the parameter to which heterogeneity will be introduced, and (ii) the distribution from which the parameter values will be drawn. For the parameter, we choose $\tau$ (we find that choosing $\epsilon$ has little effect and the wave always propagates). For the distribution, we use binary heterogeneity for simplicity, where each cell has probability $\phi$ to be on with $\tau_{\rm on}>7$, and probability $1-\phi$ to be off with $\tau_{\rm off}=5$. In Appendix A, we compare binary heterogeneity to the continuous alternatives of log-normal and log-uniform heterogeneity. We find that all three choices lead to similar results, and therefore we focus only on binary heterogeneity in what follows.

\subsection{Spatial correlations}
We introduce radially directed spatial correlations governed by an order parameter $\rho$ \cite{zhai2019statistics}. With $\rho=0$, the on-cells are randomly distributed in the excitable medium. With $\rho=1$, radii of perfectly correlated on or off-cells extend from the center to the periphery of the excitable medium. To achieve a given level of correlation, we perform the following procedure. First, we randomly assign the cells in the central 6 by 6 grid to be on with probability $\phi$, and off otherwise. Then, for each cell with unassigned neighbors (in random order), we choose a neighbor at random. With probability $\rho$ we assign the neighbor to the same on/off state as the original cell. Otherwise, we assign the on/off state from the binary distribution. We repeat these steps until all cells are assigned.

\section{Results}
\subsection{Heterogeneity produces three dynamic regimes}
\begin{figure}
	\center{\includegraphics[width=0.5\textwidth]
		{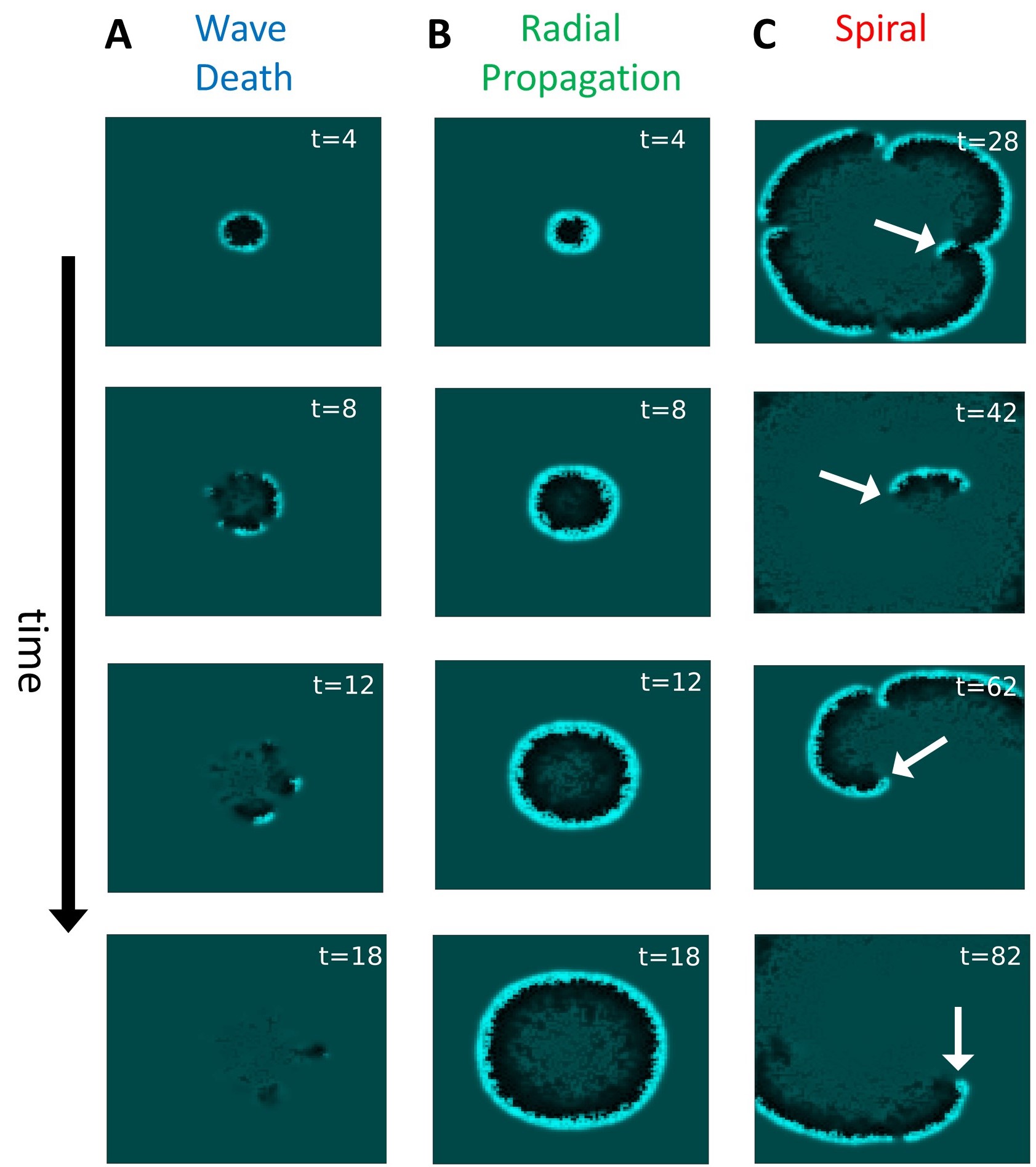}  }
	\caption{Examples of the three dynamic regimes. Brightness is proportional to signal strength $u$, showing wavefront (cyan, $u\to1$) and refractory wave back (black, $u < 0$). (A) Wave death: signal dies at $t=18$ before reaching any edge. (B) Radial propagation: wave propagates to edge and is absorbed at boundaries. (C) Spiral: wave spins back on itself and persists indefinitely; white arrow indicates spiral head.} \label{Fig2}
\end{figure}

First we focus on uncorrelated heterogeneity ($\rho=0$). As we vary the fraction of on-cells $\phi$ and their response timescale $\tau_{\rm on}$, we find three dynamic regimes, which we denote wave death, radial propagation, and spiral (Fig.\ \ref{Fig2}). Wave death means that the wave dies ($u\to0$) before reaching any edge (Fig.\ \ref{Fig2}A). Radial propagation means that the wave propagates from the center to the edge (Fig.\ \ref{Fig2}B). Spiral means that the wave forms a spiral pattern and can survive indefinitely (Fig.\ \ref{Fig2}C). We distinguish between radial propagation and spiral using the following criterion: if each cell undergoes at most one excitation we define the dynamics as radial propagation; otherwise we define the dynamics as spiral.

 \begin{figure*}
 	\center{\includegraphics[width=0.8\textwidth]
 		{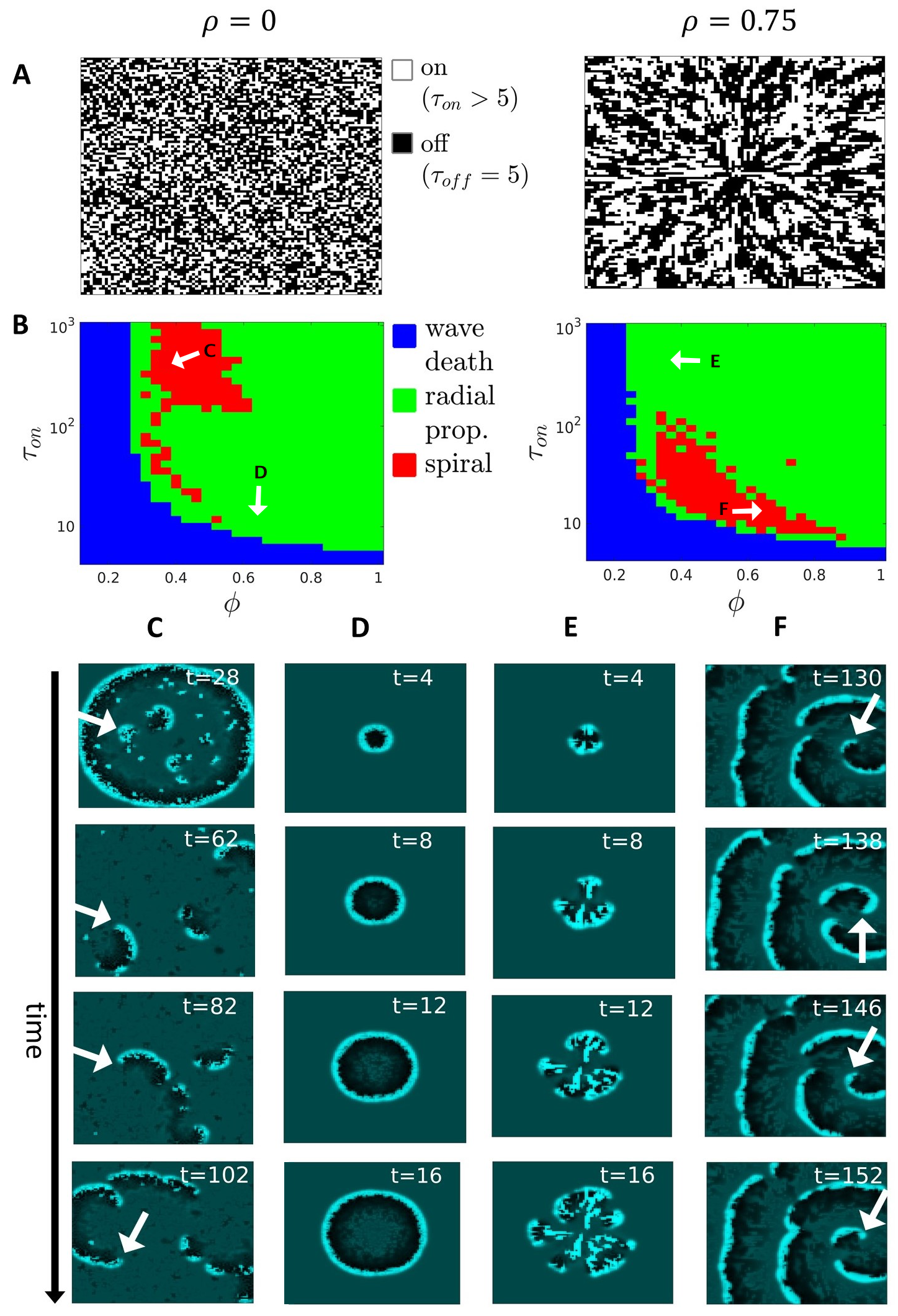}  }
 	\caption{Correlated heterogeneity can either suppress or promote spiral waves. (A) With no correlation ($\rho=0$, left), on-cells are randomly distributed in the medium. Radial correlation ($\rho=0.75$, right) produces channels of on-cells. (B) Correlation suppresses spiraling at large $\tau_{\rm on}$ (top red regime) but promotes spiraling at small $\tau_{\rm on}$ (bottom red regime). We refer to these subregimes as secondary spirals and breakup spirals, respectively. Phase diagrams in B contain 20 trials; spiral phase means that a spiral was observed in at least one trial, and otherwise we label with wave death or radial propagation, whichever is more frequent. Labels C-F in B correspond to the dynamics shown in panels C-F. White arrows in C-F indicate spiral head.} \label{Fig3}
 \end{figure*}
 
For $\rho=0$ (Fig.\ \ref{Fig3}A, left), the phase diagram of these dynamic regimes in the space of $\phi$ and $\tau_{\rm on}$ is shown in Fig.\ \ref{Fig3}B (left). We see that for sufficiently small $\phi$, there are not enough on-cells to support the excitable wave; the wave cannot transmit to the edge, resulting in wave death (blue). In contrast, for intermediate and large $\phi$ there are enough on-cells to support the wave, and we have mostly radial propagation (green). However, for a specific range of $\phi$ and when $\tau_{\rm on}$ is sufficiently large, the wave becomes long-lived, and we see that this gives rise to spiraling (red).

\subsection{Spatial correlation can either suppress or promote spiraling}
Next, we investigate the effects of spatially correlated heterogeneity ($\rho >0$). Correlation can occur, for example, if excitation properties are inherited from cell to cell as the cells grow radially outward to form the community \cite{zhai2019statistics}. Intuitively, we would expect that because high radial correlation creates channels of on-cells (Fig.\ \ref{Fig3}A, right), it should promote radial propagation and suppress spiraling. However, we find that it can either suppress or promote spiraling depending on the parameter regime. Specifically, as seen in Fig.\ \ref{Fig3}B, the top portion of the red spiral regime in the phase diagram shrinks compared to the $\rho=0$ case, as expected, whereas the bottom portion of the red spiral regime expands. The latter is a surprising feature because one expects channeling to be beneficial to radial propagation and thus detrimental to spiraling. 

Because of the different responses to correlated heterogeneity, we hypothesize that spirals in the top and bottom portions of the spiral regime are formed by different mechanisms. To investigate this hypothesis, we focus on specific examples within these regimes (see the labels C-F in Fig.\ \ref{Fig3}B, which correspond to Fig.\ \ref{Fig3}C-F). In the top regime, when the heterogeneity is uncorrelated, we see in Fig.\ \ref{Fig3}C that some cells remain on after the wave has passed. These cells then trigger secondary excitations that become spirals. We therefore refer to these spirals as secondary spirals. When the heterogeneity is correlated, we see in Fig.\ \ref{Fig3}E that these secondary spirals are suppressed by the channeling effect, and the wave propagates radially via the channels.

In the bottom regime, when the heterogeneity is uncorrelated, we see in Fig.\ \ref{Fig3}D that the wave propagates radially. However, when the heterogeneity is correlated, we see in Fig.\ \ref{Fig3}F that spirals form. The formation mechanism is different here than it is for the secondary spirals. Here, when the wavefront confronts a barrier of off-cells between channels, the wavefront breaks. Broken pieces of the wavefront then become spirals. We therefore refer to these spirals as breakup spirals. Because the channeling is what causes the breakup, breakup spirals are enhanced by correlated heterogeneity, in contrast to secondary spirals.

We can now understand why secondary spirals form at large $\tau_{\rm on}$, whereas breakup spirals form at small $\tau_{\rm on}$ (Fig.\ \ref{Fig3}B). $\tau_{\rm on}$ sets the timescale for the excitation, including the refractory period. For a secondary spiral to form, large $\tau_{\rm on}$ is necessary because this allows some of the on-cells---specifically, those that are surrounded by other on-cells---to remain excited long after the wave has passed. In contrast, for a breakup spiral to form, small $\tau_{\rm on}$ is necessary because this means that the cells near the broken piece have already completed their refractory period. They are able to become excited again, which allows propagation of the breakup spiral.

 \subsection{Dependence of spiral period on anchor size}
\begin{figure}
	\center{\includegraphics[width=0.5\textwidth]
		{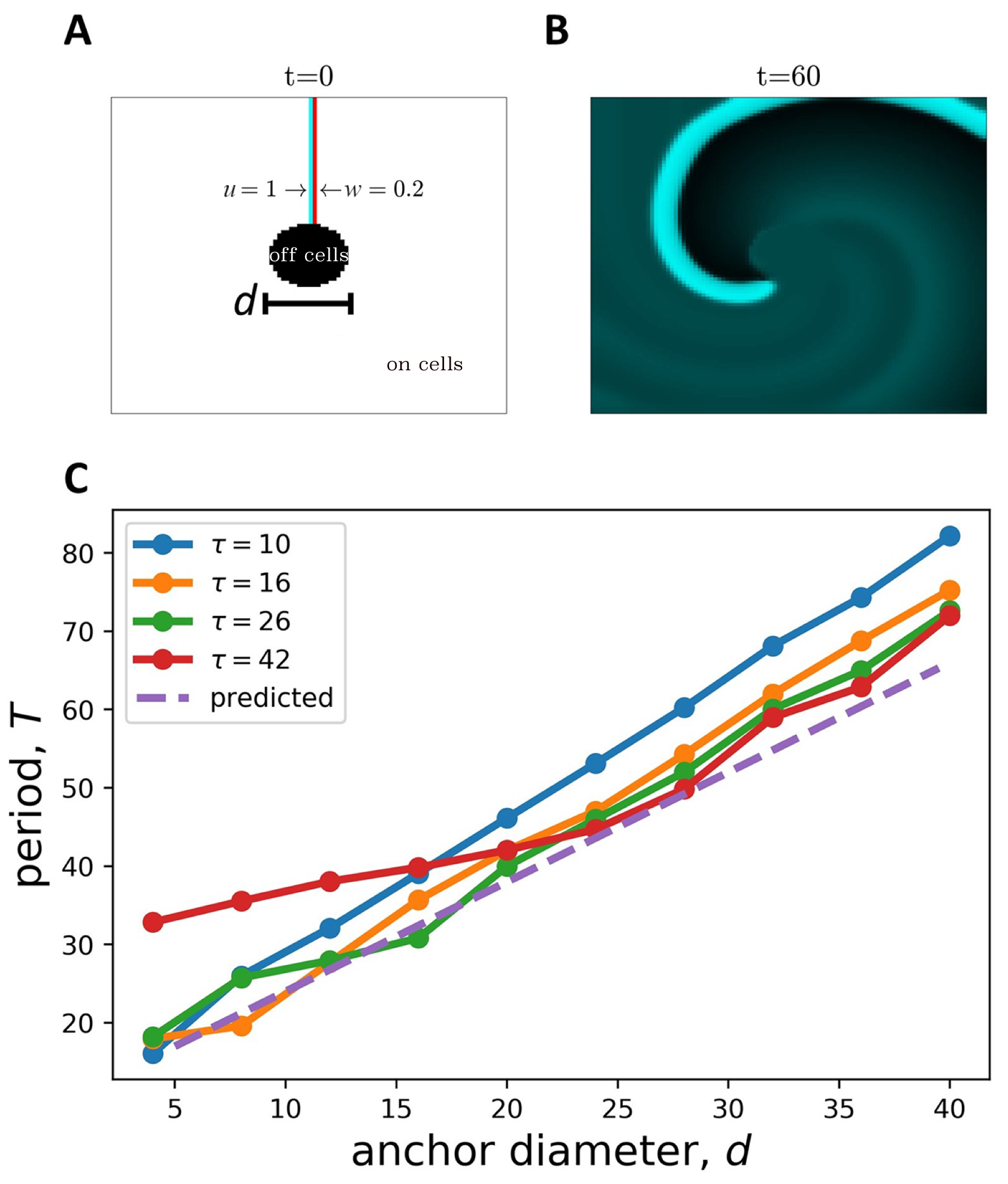}  }
	\caption{Simplified example to investigate spiral period. (A) A circular off-cell anchor (black) with diameter $d$ is surrounded by on-cells (white). Cells are initialized to trigger a counter-clockwise spiral, shown in B. (C) Period $T$ as a function of $d$ with various values of $\tau$.} \label{Fig4}
\end{figure}

Our investigation of specific examples in Fig.\ \ref{Fig3}C-F suggests that when spirals form, they circulate around clusters of off-cells in the medium, which we refer to as anchors. This raises the question of how the properties of the anchors influence the characteristics of  spiraling, e.g.\ the spiral period. To address this question, we first consider a simplified example shown in Fig.\ \ref{Fig4}A. In this example, a circular off-cell anchor with diameter $d$ is surrounded by on-cells. The two central columns of cells above the anchor are initiated with $u=1$ to trigger the wave, and $w=0.2$ (a refractory state) to prevent wave propagation to the right, respectively. As a result, a spiral forms that travels around the anchor in a counter-clockwise direction (Fig.\ \ref{Fig4}B).

In a homogeneous medium, the speed of a wave propagating according to Eqs.\ \ref{rd1} and \ref{rd2} is known \cite{Tuckwell} in the limit of $\tau\to\infty$ to be $v = \sqrt{\epsilon/2}$. Because the distance that the spiral travels around the anchor is $\pi d$, the period of rotation is $T = \pi d/v$, or
\begin{equation}
\label{T}
T=\frac{\pi d}{\sqrt{\epsilon/2}}.
\end{equation}
Fig.\ \ref{Fig4}C shows a test of this prediction in our simple example. We see that the measured period agrees well with the predicted period, and that the agreement improves as $\tau$ increases, as expected. Furthermore, we see a deviation from the prediction for small $d$ when $\tau$ is large, which can be understood as follows. When the anchor size is small enough that the period is less than the recovery time $\tau$, the wavefront is disrupted by the previous wave back, and the period is increased. Specifically, this condition will occur when the predicted period in Eq.\ \ref{T} is larger than $\tau$, which sets a value of $d > \tau\sqrt{\epsilon/2}/\pi$ below which the prediction should fail. For example, with $\tau = 42$ and $\epsilon = 10$, we expect a deviation below $d \approx 30$, which is consistent with Fig.\ \ref{Fig4}C (red curve).

\subsection{Dependence of spiral period on heterogeneity}
In the heterogeneous system, the anchors consist of the naturally occurring clusters of off-cells. The spatial statistics of these clusters are described by percolation theory \cite{Stauffer}. Therefore, the goal of this section is to use percolation theory to investigate how the heterogeneity parameters $\phi$ and $\rho$ affect the spatial structure of the community and in turn, the spiral dynamics.

Intuitively we expect that at any $\rho$, the spiral period is a monotonically decreasing function of $\phi$ because the off-cell clusters shrink with $\phi$. To make this expectation quantitative, we turn to percolation theory. Two-dimensional percolation theory on a square lattice applies when $\rho = 0$ and predicts a critical threshold $\phi_c \approx 0.59$, above which there exists a giant on-cell cluster spanning the medium. We expect this regime to be dominated by radial propagation. Conversely, below a complementary threshold of $\phi = 1-\phi_c \equiv p_c \approx 0.41$, there exists a giant off-cell cluster. We expect this regime to be dominated by wave death. Indeed, we see in Fig.\ \ref{Fig3}B (left) that, roughly speaking, radial propagation occurs for $\phi > \phi_c$, wave death occurs for $\phi < p_c$, and spirals occur within $p_c < \phi < \phi_c$. These regimes are rough because percolation theory accounts for the static structure of the lattice but not the excitable dynamics.

\begin{figure*}
	\center{\includegraphics[width=1.0\textwidth]
		{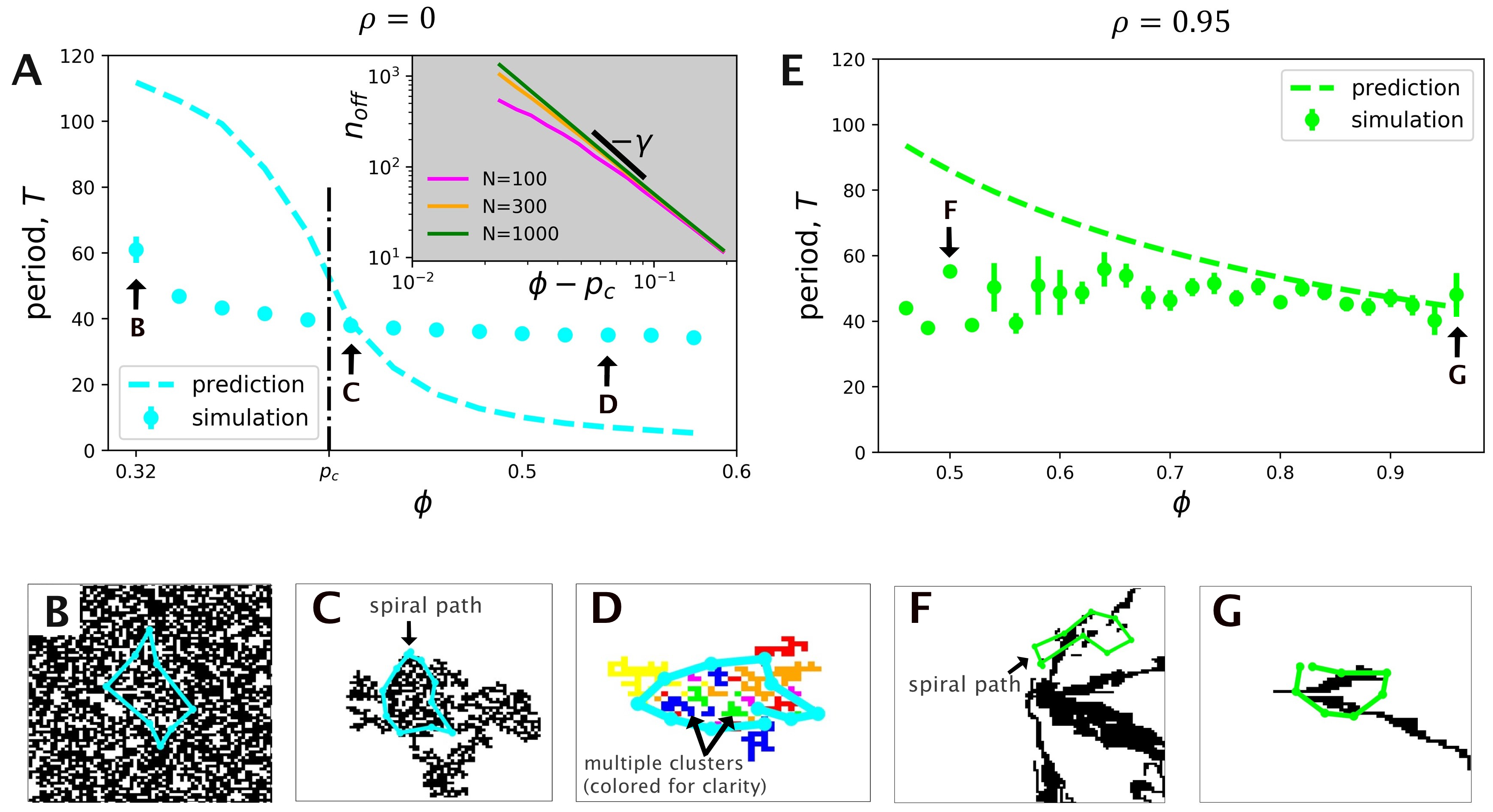}  }
	\caption{Dependence of spiral period on heterogeneity parameters with predictions from percolation theory. (A-D) Uncorrelated heterogeneity ($\rho=0$). A shows period $T$ vs.\ fraction of on-cells $\phi$ predicted by two-dimensional percolation theory and computed from simulations. Prediction succeeds qualitatively but fails quantitively as described in text and illustrated in B-D. Inset confirms scaling prediction of percolation theory for various $N\times N$ system sizes. B-D show examples (labeled in A) of spiral dynamics (cyan, trace of spiral head) and anchoring cluster(s) of off-cells (in D there are many clusters, distinguished by color). (E-G) Strongly correlated heterogeneity ($\rho = 0.95$). E is as in A but with one-dimensional percolation theory to describe quasi-one-dimensional clusters. Prediction succeeds at large $\phi$. F and G are as in B-D. In A-D, $\tau_{\rm on} = 240$; in E-F, $\tau_{\rm on} = 9$. In A and E, period is calculated as average time between excitations ($u > 0.6$) from cases out of 1000 trials where spiraling occurred; error bars are standard error.} \label{Fig5}
\end{figure*}

Near either percolation threshold, percolation theory provides scaling relations for the cluster properties. Therefore, we use percolation theory to understand how the anchor size scales with $\phi$. Specifically, above $p_c$, the mean off-cell cluster size scales as \cite{Stauffer}
\begin{equation}
n_{\rm off} \sim (\phi- p_c)^{-\gamma}
\end{equation}
where $\gamma=43/18\approx 2.39$. We confirm this scaling with progressively larger lattice sizes in the inset of Fig.\ \ref{Fig5}A. We see that the mean anchor size is indeed a decreasing function of $\phi$.

The period should depend on the circumference of the off-cell cluster being circulated. To the extent that the fractal dimension \cite{Stauffer} of the off-cell cluster is near two (at $\phi = p_c$ it is $91/48\approx 1.90$), the mean circumference should be roughly $\pi\sqrt{n_{\rm off}}$. Therefore, following Eq.\ \ref{T}, we predict that the spiral period should be
\begin{equation}
T=\frac{\pi \sqrt{n_{\rm off}}}{\sqrt{\epsilon/2} }.
\end{equation}
This prediction is shown in Fig. \ref{Fig5}A (cyan dashed line), where we measure $n_{\rm off}$ from the lattices. Above $\phi = p_c$ we see the power-law decay from percolation theory, while below $\phi = p_c$ the divergence of $n_{\rm off}$ and thus $T$ is prevented by the finite size of the lattice. We then compare this prediction to the dynamic simulation results by plotting the average spiral period (cyan data points in Fig. \ref{Fig5}A). We see for both the prediction and the simulation results that $T$ decreases with $\phi$ as expected, but that otherwise the agreement is poor.

To understand the deviation between the prediction and the simulations, we investigate the dynamics of individual spirals at various values of $\phi$. When $\phi< p_c$ (Fig. \ref{Fig5}B), the giant off-cell cluster spans the whole system. In contrast, the spiral typically circulates only a small part of the cluster. Therefore the observed period is less than predicted, as seen in Fig.\ \ref{Fig5}A. When $\phi \approx p_c$ (Fig. \ref{Fig5}C), the spiral typically circulates the bulk of a single off-cell cluster. The spiral avoids the fractal ``arms'' of the cluster, but the cluster is also not maximally dense. These two effects compensate, such that the predicted and observed periods generally agree. When $\phi > p_c$ (Fig. \ref{Fig5}D), the mean off-cell cluster size is small, but the spiral typically circulates multiple nearby clusters. Therefore the observed period is greater than predicted, as seen in Fig.\ \ref{Fig5}A.

Thus, we conclude that with uncorrelated heterogeneity ($\rho = 0$), percolation theory provides qualitative intuition for the dependence of the spiral period on the on-cell fraction, but not quantitative predictions due to the fact that it cannot capture the details of the propagation dynamics. What about highly correlated heterogeneity, when $\rho$ is large?

Standard percolation theory assumes $\rho = 0$ and therefore does not apply for general $\rho$. However, when $\rho$ is very large ($\rho\to1$), the off-cell clusters become highly distended in the radial direction due to the correlation. Therefore, they become quasi-one-dimensional, and we may expect that tools from one-dimensional percolation theory \cite{Stauffer} could provide a predictive understanding. Indeed, if we consider any radius of the system as a one-dimensional lattice, we can calculate, for a randomly chosen off-cell in that lattice, the mean length of the chain of off-cells of which it is a member. The result, derived in Appendix B, is
\begin{equation}
\label{n1D}
n_{\rm off}^{\rm 1D} = \frac{2}{(1-\rho)}\frac{1}{\phi}-1,
\end{equation}
The circumference of the chain (ignoring the ends) is twice this length, such that a spiral circulating with speed $\sqrt{\epsilon/2}$ should have a period
\begin{equation}
\label{T1D}
T = \frac{2}{\sqrt{\epsilon/2}}\left[\frac{2}{(1-\rho)}\frac{1}{\phi}-1\right].
\end{equation}
When $\phi$ is small, there are many off-cells in the system. Chains of off-cells from neighboring radii will fall next to each other, and the quasi-one-dimensional nature of the clusters will be lost. Therefore, we expect Eq.\ \ref{T1D} to fail for small $\phi$. On the other hand, for large $\phi$, chains of off-cells will be rare, and we expect Eq.\ \ref{T1D} to hold.

The prediction in Eq.\ \ref{T1D} is compared with simulations in Fig.\ \ref{Fig5}E for $\rho = 0.95$. As expected, we see that the prediction fails at small $\phi$ but succeeds at large $\phi$. Indeed, as seen in Fig.\ \ref{Fig5}F, at small $\phi$, chains of off-cells from neighboring radii fall next to each other, and the clusters are no longer quasi-one-dimensional. In contrast, as seen in Fig.\ \ref{Fig5}G, at large $\phi$, the clusters quasi-one-dimensional and well separated, and the spiral generally circulates one of the clusters.

\section{Discussion}
We have developed a model that describes wave propagation in a spatially correlated heterogeneous medium. Despite the minimal nature of the model, and the idealization of cells as squares on a regular lattice, our model supports multiple dynamic regimes including wave death, radial propagation, and spiraling. We have discovered that spirals arise via two distinct mechanisms---secondary triggers and wavefront breakup---and that spatial correlations in the heterogeneity suppress the former and promote the latter, in two distinct regimes within parameter space. Furthermore, we have shown using percolation theory that the structural properties of the heterogeneity influence the dynamic properties of spiraling (the period), and that the predictive power of percolation theory is best when the spatial correlation is strong. Taken together, our results suggest that the spatial structure of cell-to-cell heterogeneity can have important consequences for directed signal propagation in cellular communities.

The minimal nature of the model facilitates its application as a phenomenological description of experiments on signaling in cellular communities. Because there are only two parameters describing the dynamics (the time scale and the excitation strength), and two parameters describing the heterogeneity (the fraction and spatial correlation of on-cells), these parameters could be easily calibrated to experimental observations. Because the parameters and variables are dimensionless, the cell length scale and signal time scale would set the units of length and time, and then other observables such as the wavelength  and fraction of signaling cells could calibrate other parameters to the point where predictions could be made and tested with no further free parameters \cite{JosephCellSystem}.

Our work could also provide insights on how one might engineer heterogeneous media to produce spiral waves. First, the excitability threshold must be low; in this work we set it to zero (in the typical FN there is an additional threshold parameter in Eq.\ \ref{rd1}). Second, an intermediate degree of heterogeneity is required, with not too many or too few excitable cells (Fig.\ \ref{Fig3}B), which could be controlled by genetic engineering \cite{JosephCellSystem} or different fractions of a multi-species cell mixture. Third, either a large or a small excitation timescale is required depending on the degree of spatial correlation (Fig.\ \ref{Fig3}B). The excitation timescale could be changed by mutation \cite{JosephCellSystem}, and the spatial correlation could be controlled by modulating the degree of cell redistribution in a microfludic device or by using strains with differential growth rates.

Our model has limitations, and several extensions are natural. Although we check that our results are insensitive to the heterogeneity distribution (Appendix A), we only add heterogeneity to one parameter (the timescale). It may be more realistic to add heterogeneity to all parameters \cite{potter2016communication}, including the excitation strength and, if included, the excitation threshold. Moreover, we assume that cells exist on a regular square lattice and that heterogeneity is only correlated in the radial direction. It may be more realistic to add disorder to either the lattice structure, cell size \cite{JosephCellSystem}, or spatial direction of the heterogeneity \cite{zhai2019statistics}. It will be interesting to see how our model can be generalized, as well as how it can be used to understand or engineer experimental systems in future work.

\appendix

\section{Robustness to heterogeneity distirbution}

\begin{figure}
	\center{\includegraphics[width=0.4\textwidth]
		{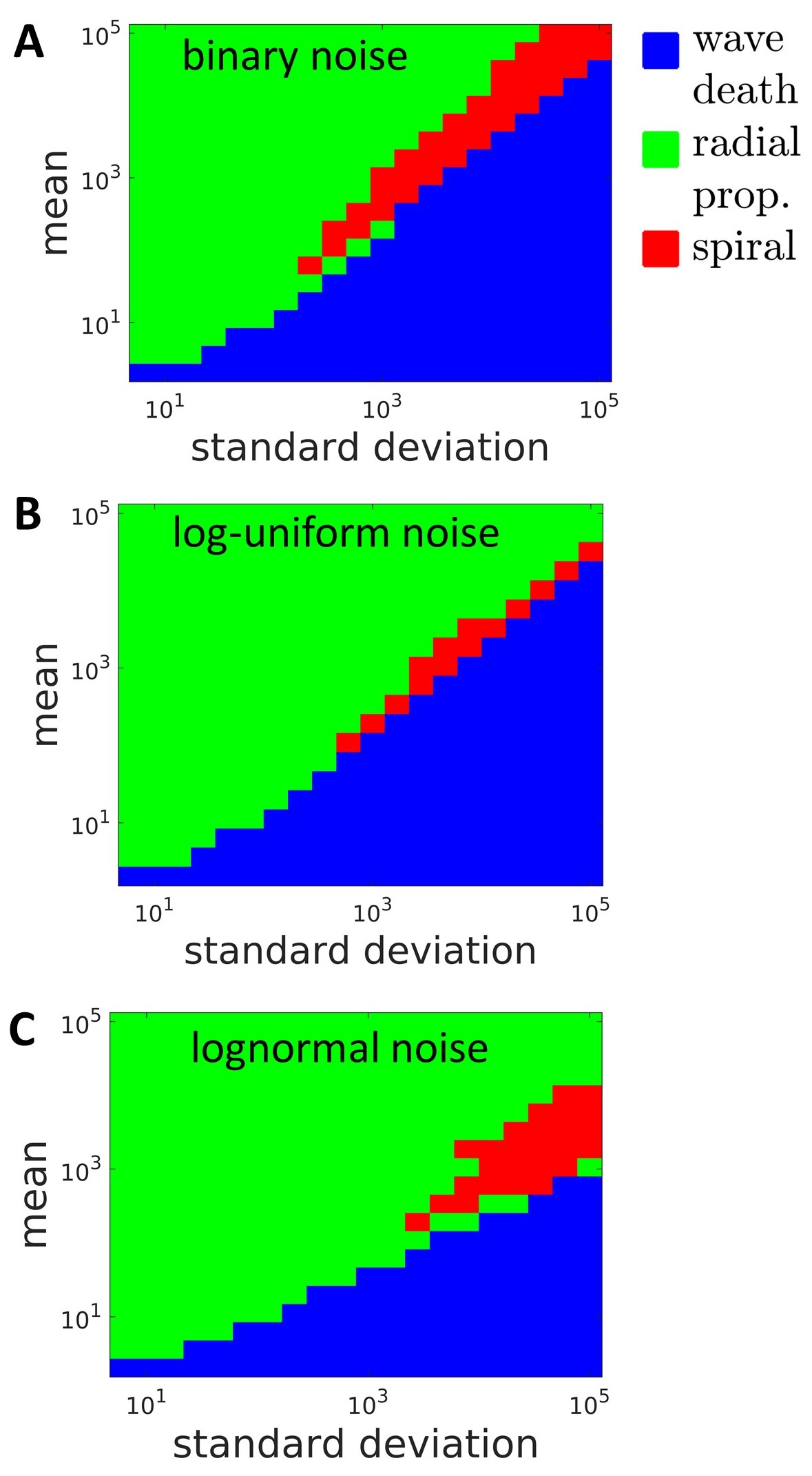}  }
	\caption{Phase diagrams in the parameter space of mean and standard deviation with three different choices for the distribution of $\tau$ values: (A) binary, (B) log-uniform, and (C) lognormal. We see that the structure of the phase diagram is largely insensitive to the choice of distribution.} \label{Fig6}
\end{figure}

In the main text we use a binary distribution for heterogeneity in the parameter $\tau$: a fraction $\phi$ of cells have timescale $\tau_{\rm on}$ while the rest have $\tau_{\rm off} = 5$. Here we compare this distribution to two others, a log-uniform and a lognormal distribution (both ensure that $\tau$ stays positive), in terms of the resulting wave dynamics. To compare all distributions on equal footing, we plot the dynamic phases in the space of the distribution's mean and standard deviation. We see in Fig.\ \ref{Fig6} that the phase diagram of wave death (blue), radial propagation (green), and spirals (red) are qualitatively similar in all three cases. We conclude that the relationship between wave dynamics and heterogeneity is largely independent of the distribution from which the heterogeneity is drawn in our model.

\section{Mean length of off-cell chain in 1D}

Consider any radius of the system as a one-dimensional lattice of $L$ cells. Denote the center and edge of the radius as the left and right end of the lattice, respectively. Calling any cell in this lattice a ``mother'' cell $m$, and its rightward neighbor a ``daughter'' cell $d$, the conditional probabilities $P(d|m)$ of the state of the daughter given the state of the mother are
\begin{align}
\label{p1}
P(d={\rm on}|m={\rm on})&=\rho+(1-\rho)\phi, \\
\label{p2}
P(d={\rm off}|m={\rm off})&=\rho+(1-\rho)(1-\phi), \\
\label{p3}
P(d={\rm on}|m={\rm off})&=(1-\rho)\phi, \\
\label{p4}
P(d={\rm off}|m={\rm on})&=(1-\rho)(1-\phi).
\end{align}
In Eqs.\ \ref{p1} and \ref{p2}, the daughter is guaranteed to be in the same state as the mother with probability $\rho$ (first term), and with probability $1-\rho$ it could also be in the same state by chance, either on with probability $\phi$ or off with probability $1-\phi$, respectively (second term). In Eqs.\ \ref{p3} and \ref{p4}, the daughter is in the opposite state as the mother, which requires both anti-correlation with probability $(1-\rho)$ and selection of the opposite state by chance with probability $\phi$ or $1-\phi$, respectively.

Given these conditional probabilities, we calculate the probability $q_s$ that a randomly chosen cell is at the left end of a chain of off-cells of size $s$. It must be off with probability $1-\phi$, its mother must be on with probability $P(m={\rm on}|d={\rm off})$, the next $s-1$ rightward cells must be off with probability $P(d={\rm off}|m={\rm off})$, and the next cell after that must be on with probability $P(d={\rm on}|m={\rm off})$, giving
\begin{align}
q_s =\ &(1-\phi)P(m={\rm on}|d={\rm off}) \nonumber \\
	&\times P(d={\rm off}|m={\rm off})^{s-1} \nonumber \\
	&\times P(d={\rm on}|m={\rm off}) \\
\label{q}
=\ &\phi^2(1-\phi)(1-\rho)^2[\rho+(1-\rho)(1-\phi)]^{s-1}.
\end{align}
The second step follows from Eqs.\ \ref{p2} and \ref{p3} and Bayes' theorem,
\begin{align}
P(m={\rm on}|d={\rm off}) =\ &\frac{P(m={\rm on})P(d={\rm off}|m={\rm on})}{P(d={\rm off})} \nonumber\\
=\ &\frac{\phi(1-\rho)(1-\phi)}{1-\phi} = \phi(1-\rho),
\end{align}
where we have used Eq.\ \ref{p4}.

The probability $q_s$ is equivalent to the number $N_s$ of chains of size $s$, divided by the lattice size $L$. Similarly, the probability that a randomly chosen cell is a member of a chain of size $s$ (not just the left end) is $N_ss/L$. Finally, the probability $P_s$ that a randomly chosen cell is a member of a chain of size $s$, given that it is an off-cell, is $N_ss/L$ divided by the probability of being an off-cell, $1-\phi$. Therefore, we have
\begin{equation}
P_s = \frac{q_ss}{1-\phi}.
\end{equation}
Thus, the mean length of a chain containing a randomly chosen off-cell is
\begin{equation}
n_{\rm off}^{\rm 1D} = \sum_{s = 1}^\infty P_s s = \frac{1}{1-\phi}\sum_{s = 1}^{\infty} q_s s^2 =
\frac{2}{(1-\rho)}\frac{1}{\phi}-1,
\end{equation}
as in Eq.\ \ref{n1D}, where we have used Eq.\ \ref{q} and the properties of a geometric series. Note that the mean chain size decreases with $\phi$ and increases with $\rho$ until diverging for $\rho \to 1$, as expected.

\acknowledgments
This work was supported by the National Institute of General Medical Sciences (R01 GM121888 to G.M.S.\ and A.M.) and the Simons Foundation (376198 to A.M.).


\begin{thebibliography}{35}

\bibitem{BalleriniPNAS}
Ballerini M, Cabibbo N, Candelier R, Cavagna A, Cisbani E, Giardina I, Lecomte
  V, Orlandi A, Parisi G, Procaccini A (2008) Interaction ruling animal
  collective behavior depends on topological rather than metric distance:
  evidence from a field study.
\newblock {\em Proceedings of the National Academy of Sciences of the United
  States of America} 105(4):1232--1237.

\bibitem{JacobAP}
Ben-Jacob E, Cohen I, Levine H (2000) Cooperative self-organization of
  microorganisms.
\newblock {\em Advances in Physics} 49(4):395--554.

\bibitem{FlynnPRE}
Flynn MR, Kasimov AR, Nave JC, Rosales RR, Seibold B (2009) Self-sustained
  nonlinear waves in traffic flow in {\em Physical Review E}.
\newblock p. 056113.

\bibitem{DaninoNature}
Danino T, Mondragónpalomino O, Tsimring L, Hasty J (2010) A synchronized
  quorum of genetic clocks.
\newblock {\em Nature} 463(7279):326--30.

\bibitem{SchellingJMS}
Schelling TC (1971) Dynamic models of segregation.
\newblock {\em Journal of Mathematical Sociology} 1(2):143--186.

\bibitem{NoorbakhshPRE}
Noorbakhsh J, Schwab DJ, Sgro AE, Gregor T, Mehta P (2015) Modeling
  oscillations and spiral waves in dictyostelium populations.
\newblock {\em Physical Review E Statistical Nonlinear \& Soft Matter Physics}
  91(6):062711.

\bibitem{SchellersheimWIR}
Meier-Schellersheim M (2009) Multi-scale modeling in cell biology.
\newblock {\em Wiley Interdiscip Rev Syst Biol Med} 1(1):: 4–14.

\bibitem{QutubEMBM}
Qutub AA, Mac GF, Karagiannis ED, Vempati P, Popel AS (2009) Multiscale models
  of angiogenesis.
\newblock {\em IEEE Engineering in Medicine \& Biology Magazine the Quarterly
  Magazine of the Engineering in Medicine \& Biology Society} 28(2):14--31.

\bibitem{SouthernPBMB}
Southern J, Pitt-Francis J, Whiteley J, Stokeley D, Kobashi H, Nobes R, Kadooka
  Y, Gavaghan D (2008) Multi-scale computational modelling in biology and
  physiology.
\newblock {\em Progress in Biophysics and Molecular Biology} 96(1):60 -- 89.
\newblock Cardiovascular Physiome.

\bibitem{LilienkampPRL}
Lilienkamp T, Christoph J, Parlitz U (2017) Features of chaotic transients in
  excitable media governed by spiral and scroll waves.
\newblock {\em Phys.rev.lett} 119(5).

\bibitem{PrindleNature}
Prindle A, Liu J, Asally M, Ly S, Garcia-Ojalvo J, Süel GM (2015) Ion channels
  enable electrical communication in bacterial communities.
\newblock {\em nature} 527:59–63.

\bibitem{LiuNature}
Liu J, Prindle A, Humphries J, Gabalda-Sagarra M, Asally M, yeon D.~Lee D, Ly
  S, Garcia-Ojalvo J, Süel GM (2015) Metabolic co-dependence gives rise to
  collective oscillations within biofilms.
\newblock {\em nature} 523:550--554.

\bibitem{BenningerBioPhysJ}
Benninger RKP, Zhang M, Head WS, Satin LS, Piston DW (2008) Gap junction
  coupling and calcium waves in the pancreatic islet.
\newblock {\em Biophysical Journal} 95(11):5048--5061.

\bibitem{JosephCellSystem}
Larkin JW, Zhai X, Kikuchi K, Redford SE, Prindle A, Liu J, Greenfield S,
  Walczak AM, Garcia-Ojalvo J, Mugler A, S\"uel GM (2018) Signal percolation
  within a bacterial community.
\newblock {\em Cell systems} 7(2):137--145.

\bibitem{zhai2019statistics}
Zhai X, Larkin JW, Kikuchi K, Redford SE, S\"uel GM, Mugler A (2019) Statistics
  of correlated percolation in a bacterial community.
\newblock {\em arXiv preprint arXiv:1906.06450}.

\bibitem{BubChaos}
Bub G, Shrier A (2002) Propagation through heterogeneous substrates in simple
  excitable media models.
\newblock {\em Chaos} 12(12):747--753.

\bibitem{KinoshitaPRE}
Kinoshita S, Iwamoto M, Tateishi K, Suematsu NJ, Ueyama D (2013) Mechanism of
  spiral formation in heterogeneous discretized excitable media.
\newblock {\em Physical Review E Statistical Nonlinear \& Soft Matter Physics}
  87(6):062815.

\bibitem{UzzamanCR}
Uzzaman M, Honjo H, Takagishi Y, Emdad L, Magee AI, Severs NJ, Kodama I (2000)
  Remodeling of gap junctional coupling in hypertrophied right ventricles of
  rats with monocrotaline-induced pulmonary hypertension.
\newblock {\em Circulation Research} 86(8):871--878.

\bibitem{BubPRL}
Bub G, Shrier A, Glass L (2002) Spiral wave generation in heterogeneous
  excitable media.
\newblock {\em Physical Review Letters} 88(5):058101.

\bibitem{SteinbockScience}
Steinbock O, Kettunen P, Showalter K (1995) Anisotropy and spiral organizing
  centers in patterned excitable media.
\newblock {\em Science} 269(5232):1857--1860.

\bibitem{TinsleyPCCP}
Tinsley MR, Taylor AF, Huang Z, Showalter K (2011) Complex organizing centers
  in groups of oscillatory particles.
\newblock {\em Physical Chemistry Chemical Physics Pccp} 13(39):17802--17808.

\bibitem{TothPRE}
Toth R, De LCB, Stone C, Masere J, Adamatzky A, Bull L (2009) Spiral formation
  and degeneration in heterogeneous excitable media.
\newblock {\em Physical Review E Statistical Nonlinear \& Soft Matter Physics}
  79(2):035101.

\bibitem{GesteiraPD}
Gómez-Gesteira M, Fernández-García G, Muñuzuri AP, Pérez-Muñuzuri V,
  Krinsky VI, Starmer CF, Pérez-Villar V (1994) Vulnerability in excitable
  belousov-zhabotinsky medium: from 1d to 2d.
\newblock {\em Physica D Nonlinear Phenomena} 76(4):359--368.

\bibitem{AlievCSF}
Aliev RR (1995) Heart tissue simulations by means of chemical excitable media.
\newblock {\em Chaos Solitons \& Fractals} 5(3–4):567--574.

\bibitem{PertsovCirR}
Pertsov AM, Davidenko JM, Salomonsz R, Baxter WT, Jalife J (1993) Spiral waves
  of excitation underlie reentrant activity in isolated cardiac muscle.
\newblock {\em Circulation research} 72(3):631--650.

\bibitem{ZimikArXiv}
Zimik S, Majumder R, Pandit R (2018) Ionic-heterogeneity-induced spiral-and
  scroll-wave turbulence in mathematical models of cardiac tissue.
\newblock {\em arXiv preprint arXiv:1807.04546}.

\bibitem{KastbergerPO}
Kastberger G, Schmelzer E, Kranner I (2008) Social waves in giant honeybees
  repel hornets.
\newblock {\em PLoS ONE} 3(9):e3141.

\bibitem{FitzHugh}
FitzHugh R (1961) Impulses and physiological states in theoretical models of
  nerve membrane.
\newblock {\em Biophysical Journal} 1(6):445--466.

\bibitem{Nagumo}
Nagumo J, Arimoto S, Yoshizawa S (1962) An active pulse transmission line
  simulating nerve axon.
\newblock {\em Proceedings of the Ire} 50(10):2061--2070.

\bibitem{HodgkinJP}
Hodgkin AL, Huxley AF (1952) Currents carried by sodium and potassium ions
  through the membrane of the giant axon of loligo.
\newblock {\em Journal of Physiology} 116(4):449.

\bibitem{OsmanISRN}
Osman GM, Toshiyuki O (2014) Alternans and spiral breakup in an excitable
  reaction-diffusion system: A simulation study:.
\newblock {\em International Scholarly Research Notices,2014,(2014-11-12)}
  2014(3-4):14 pages.

\bibitem{Tuckwell}
Tuckwell H (1988) {\em Introduction to theoretical neurobiology}.
\newblock (the Press Syndicate of the University of Cambridge, The Pitt
  Building, Trumpington Street, Cambridge CB2 1 RP).

\bibitem{Mirollo}
Mirollo RE, Strogatz SH (1990) Synchronization of pulse-coupled biological
  oscillators.
\newblock {\em SIAM Journal on Applied Mathematics} 50(6):1645--1662.

\bibitem{Stauffer}
D S, A A (1994) {\em Introduction to percolation theory}.
\newblock (CRC Press, Boca Raton, FL).

\bibitem{potter2016communication}
Potter GD, Byrd TA, Mugler A, Sun B (2016) Communication shapes sensory
  response in multicellular networks.
\newblock {\em Proceedings of the National Academy of Sciences}
  113(37):10334--10339.

\end{thebibliography}

\end{document}